\documentclass[12pt]{JHEP3}

\usepackage{amsmath}

\title{An Exact Bosonization Rule for $c=1$ Noncritical String
 Theory}

\author{Nobuyuki Ishibashi\\
Institute of Physics, University of Tsukuba, Tsukuba, Ibaraki 305-8571, Japan\\ 
E-maile: \email{ishibash@het.ph.tsukuba.ac.jp}}

\author{Atsushi Yamaguchi\\
High Energy Accelerator Research Organization (KEK), Tsukuba, Ibaraki
305-0801, Japan\\
E-maile: \email{ayamagu@post.kek.jp}}

\abstract{We construct a string field theory for $c=1$ noncritical strings 
using the loop variables as the string field. 
We show how one can express the nonrelativistic free fermions which 
describes the theory, in terms of these string fields. }

\keywords{D-branes, Matrix Models}

\begin{document}

\section{Introduction}
Noncritical string theories provide useful toy models to study various dynamical 
aspects of string theory  (for reviews, see \cite{rev1}-\cite{rev4}). 
They are exactly solvable and have many features in common with the critical 
ones.

The $c=1$ case\footnote{See \cite{rev2}\cite{rev3} for earlier reviews. See \cite{rev4} and
references therein for recent developments including
\cite{McGreevy:2003kb}-\cite{Takayanagi:2003sm}.}
is the most interesting one in which
the two-dimensional target space
interpretation is possible.
It is solved by the matrix model techniques and can be reduced to a free nonrelativistic 
fermion system. 
On the other hand, $c=1$ noncritical string theory can be described by a string field 
theory \cite{Das-Jevicki}. 
Although the string field theory looks not so simple, the Hamiltonian can be diagonalized 
by using the free fermions. 
Therefore from the point of view of the string field theory, it will be an intriguing 
problem to find a way to express the fermions in terms of the string fields.  

There have been many attempts to give such a "bosonization rule"
\footnote{"Fermionization rule" may be a more appropriate word to be used here.}
\cite{Gross-Klebanov}-\cite{Alexandrov:2004cg}.
In \cite{Gross-Klebanov}, in particular, the fermions were
expressed approximately by using the perturbative fluctuation 
of the string field.
This is based on a standard bosonization formula for the
two-dimensional relativistic fermion.  
In this paper, we would like to propose an exact bosonization rule. 
We first rewrite the string field theory for $c=1$ strings using the loop variables. 
The Hamiltonian consists of the joining-splitting type terms, whose forms are quite 
similar to those in the string field theory for critical strings. 
Then we will give a bosonization rule to express the fermions in terms of these 
string fields. 
The bosonization rule we propose is exact even before the continuum limit is taken. 

Our bosonization rule is a generalization of the D-instanton operator proposed in 
\cite{Hanada:2004im}. 
In \cite{Hanada:2004im}, 
the D-instanton
operator was used to investigate the functional form 
of the chemical potential of D-instantons 
\cite{Hanada:2004im}-\cite{Kuroki:2007an}. 
Therefore our bosonization rule 
will be useful in understanding the nature of D-branes in $c=1$ string theory.  
We will perform some perturbative calculations using this rule. 

The organization of this paper is as follows. 
In section 2, we describe a string field theory for $c=1$ strings 
in terms of the loop variables and show that it is equivalent to the 
Das-Jevicki's formulation \cite{Das-Jevicki}. 
In section 3, we give the bosonization rule. 
In section 4, we take the double scaling limit and describe the 
string field theory and the bosonization rule in the continuum limit. 
In section 5, we describe how one can perform perturbative calculations 
using the string field theory and the bosonization rule. 
Section 6 is devoted to discussions. 

\section{Collective field theory}
$c=1$ noncritical string theory can be described by the matrix quantum mechanics:
\begin{equation}
\int\! dM \exp 
\left[
i\beta \!\!\int \!dt~ \mbox{Tr} 
\left(\frac{1}{2}\dot{M}^2-U(M)\right)
\right].
\end{equation}
Here $M(t)$ is an $N\times N$ hermitian matrix. 
We can take the double scaling limit in which $\beta\rightarrow\infty$ with an appropriate 
matrix potential $U(M)$.\footnote{This procedure involves the limit $N\to \infty$
implicitly.} 
The matrix Hamiltonian is derived from the action as  
\begin{equation}
H=
\mbox{Tr}
\left[
-\frac{1}{2\beta^2}\left(\frac{\partial}{\partial M}\right)^2
+U(M)
\right].
\label{mmhamiltonian}
\end{equation}

In order to study the $c=1$ theory, we are mainly interested in the so-called singlet sector. 
Then what is relevant is the eigenvalues $\zeta_i (i=1,\cdots ,N)$ of the matrix $M$ and 
the wave function $\Psi (\vec{\zeta})$ is given as a function of these eigenvalues. 
The Hamiltonian is expressed as a differential operator in terms of $\zeta_i$ as 
\begin{equation}
\sum_i
\left[
-\frac{1}{2\beta^2}
\triangle(\vec{\zeta})^{-1}
\left(
\frac{\partial}{\partial \zeta_i}
\right)^2
\triangle(\vec{\zeta})
+
U(\zeta_i)
\right],
\label{eigenhamiltonian}
\end{equation}
where $\triangle (\vec{\zeta})$ is the Van-der-Monde determinant.

\subsection{Loop variables}
Now we would like to construct the collective field theory for 
this matrix quantum mechanics.\footnote{A similar construction was considered in \cite{Jevicki:1991yi}.}
The basic idea of the collective field theory is to 
express the wave function of the system as a functional of the loop variable 
\begin{equation}
\varphi (l)\equiv \mbox{Tr}e^{lM}.
\end{equation}
In the matrix model, this quantity corresponds to a boundary on the worldsheet 
with length $l$. Thus we will consider this operator for $l>0$. 
The wave function $\Psi (\vec{\zeta })$ is now expressed as a functional 
$\Psi [\varphi ]$. 
As we will see, any function $\Psi (\vec{\zeta })$ can be expressed as such a 
functional, but the space of such functionals are bigger than the Hilbert space of 
the matrix quantum mechanics. 
We define the operator $\hat{\varphi}(l), \hat{\bar{\varphi}}(l)$ to be the ones 
which act as 
\begin{eqnarray}
& &
\hat{\varphi}(l)\Psi [\varphi ]
=
\varphi (l)\Psi [\varphi ],
\nonumber
\\
& &
\hat{\bar{\varphi}}(l)\Psi [\varphi ]
=
l\frac{\delta}{\delta \varphi (l)}\Psi [\varphi ],
\end{eqnarray}
on the wave function $\Psi [\varphi ]$. 
In the following, we omit the hats to represent the operators. 
$\varphi (l)$ and $\bar{\varphi}(l)$ satisfy 
\begin{equation}
[\bar{\varphi}(l),\varphi(l^\prime )]=l\delta (l-l^\prime ).
\label{commutation}
\end{equation}
It is easy to see that the Hamiltonian
eq.(\ref{mmhamiltonian}) is expressed  in terms of these variables as 
\begin{eqnarray}
H
&=&
-\frac{1}{2\beta^2}
\int dl_1dl_2
\left[
\varphi (l_1)\varphi (l_2)\bar{\varphi}(l_1+l_2)
+\varphi (l_1+l_2)\bar{\varphi}(l_1)\bar{\varphi}(l_2)
\right]
\nonumber
\\
& &
+\int dl\varphi (l)U(-\partial_l)\delta (l).
\label{collective field hamiltonian}
\end{eqnarray}

\subsection{Relation to Das-Jevicki variables}
This collective field theory is of course equivalent to the Das-Jevicki 
theory. In order to rewrite the Hamiltonian into the Das-Jevicki form, 
we express the field $\varphi (l)$ in terms of the density of the matrix 
eigenvalues $\rho (\zeta )$ as 
\begin{equation}
\varphi (l)
=
\int_{-\infty}^{\infty}d\zeta e^{\zeta l}\rho (\zeta ).
\label{rho phi}
\end{equation}
We assume that $\rho (\zeta )$ possesses a compact support on the real axis. 
The Laplace transform of $\varphi (l)$ becomes 
\begin{eqnarray}
\varphi (\zeta )
&\equiv&
\int_0^\infty dle^{-\zeta  l}\varphi (l),
\nonumber
\\
&=&
\int d\zeta^\prime 
\frac{\rho (\zeta^\prime )}{\zeta -\zeta^\prime },
\end{eqnarray}
and for $\bar{\varphi}$ we define
\begin{equation}
\bar{\varphi}(-\zeta )
\equiv
\int_0^\infty dle^{\zeta l}\bar{\varphi}(l).
\end{equation}

Then it is straightforward to show that the relation 
between our variables and the Das-Jevicki variables $\rho ,\pi$ 
is given as 
\begin{eqnarray}
\varphi (\zeta \pm i\delta )+\bar{\varphi}(-\zeta )
&=&
i\partial_\zeta \pi (\zeta )
\mp i\pi \rho (\zeta )
\nonumber
\\
&\equiv&
ip_{\mp}(\zeta ),
\label{phi p}
\end{eqnarray}
where $\zeta$ is real and $\delta >0$ is very small. 
The commutation relation (\ref{commutation}) implies that 
$\pi (\zeta )$ is the canonical conjugate of $\rho (\zeta )$. 
Using this relation, we can rewrite the Hamiltonian 
(\ref{collective field hamiltonian}) as
\begin{equation}
H
=
\int \frac{d\zeta }{2\pi}
\left[
\frac{1}{6\beta^2}(p_+^3-p_-^3)
+U(\zeta )(p_+-p_-)
\right],
\label{DJHamilton}
\end{equation}
which is exactly the Das-Jevicki Hamiltonian. 

Since 
$\rho$ and $\pi$ are defined to be hermitian operators, the Hamiltonian is hermitian. 
From eqs.(\ref{phi p}), we can obtain the hermitian conjugates 
of $\varphi$ and $\bar{\varphi}$ as 
\begin{eqnarray}
& &
\varphi^\dagger (l)=\varphi (l),
\nonumber
\\
& &
\int dle^{\zeta l}\bar{\varphi}^\dagger (l)
=
-
\left[
\int dle^{\zeta l}\bar{\varphi}(l)
  +2\mbox{Re}
\int dle^{-\zeta l}\varphi (l)
\right].
\end{eqnarray}

Before closing this subsection, one comment is in order. 
The Hamiltonian in eq.(\ref{DJHamilton}) is actually the classical part of 
the Das-Jevicki Hamiltonian. 
In Das-Jevicki's formulation\cite{Das-Jevicki}, 
there exist higher order terms, which we are not able to reproduce. 
We will come back to this point later. 

\section{Bosonization}
The string field formulation using the loop variables is 
just another expression of the familiar Das-Jevicki formalism. 
However, the loop variables are 
convenient for guessing the form of the bosonization formula. 
The algebra of the loop variables is quite analogous to that of the bosonic oscillators. 
Indeed if we compare $\varphi (l)$ and $\bar{\varphi}(l)$ to $\alpha_{n}$ and $\alpha_{-n}$ 
with $l,~(l>0)$ corresponding to $n>0$, the commutation relation eq.(\ref{commutation}) 
should correspond to 
\begin{equation}
[\alpha_n ,\alpha_{-m}]
=
n\delta_{n,m}.
\end{equation}
Using this analogy, one can guess how one can construct fermions from the bosonic operators 
$\varphi$ and $\bar{\varphi}$. 
From the usual bosonic oscillators, one can construct a fermionic operator roughly as 
\begin{equation}
:\exp \left(
-\sum_n\frac{1}{n}\alpha_nz^{-n}
\right):.
\end{equation}
Therefore it is conceivable that if we construct something like 
\begin{equation}
\exp 
\left[
- 
\int_\epsilon^\infty \frac{dl}{l}e^{-(\zeta \pm i\delta )l}\varphi (l)
\right]
\exp 
\left[
\int_\epsilon^\infty \frac{dl}{l}e^{\zeta l}\bar{\varphi }(l)
\right],
\end{equation}
it will behave as a fermionic operator. 

However things are not so straightforward. For one thing, zero modes play important roles 
in the usual bosonization and we need to find a substitute for those in the collective field theory. 
Secondly, the above analogy is not correct as to the hermiticity of the operators and it may cause 
trouble in defining the fermion conjugate to the one above. 
What we will show is that the above guess is essentially correct and we can construct fermionic 
operators in the Hilbert space of the collective field. 

\subsection{Collective field Hilbert space}
Before constructing the fermionic operators, we will construct the Hilbert space of the collective 
field $\varphi$ so that it can describe the matrix quantum mechanics. 
Let $~_\varphi \langle 0|$ be the eigenstate of $\varphi (l)$ with the eigenvalue 
$0$, i.e. 
\begin{equation}
~_\varphi \langle 0|\varphi (l)=0.
\end{equation}
For $\vec{\zeta}=(\zeta_1,\cdots ,\zeta_N)$, we define
\begin{equation}
\langle \vec{\zeta}|
\equiv 
~_\varphi\langle 0|
\exp \left[
\sum_i\int \frac{dl}{l}e^{\zeta_i l}\bar{\varphi}(l)
\right].
\end{equation}
$\langle \vec{\zeta}|$ is an eigenstate of $\varphi (l)$ and 
\begin{equation}
\langle \vec{\zeta}|\varphi (l)
=
\langle \vec{\zeta}|\sum_ie^{\zeta_i l}.
\label{eigenvalue}
\end{equation}

Now for a state $|\Psi \rangle$, we define the wave function 
$\Psi (\vec{\zeta})$ as 
\begin{equation}
\Psi (\vec{\zeta})
=
\langle \vec{\zeta}|\Psi \rangle ,
\end{equation}
and identify $\Psi (\vec{\zeta})$ with the wave function for the matrix 
eigenvalues. 
In such a representation, using 
eqs.(\ref{collective field hamiltonian})(\ref{eigenvalue}) 
we obtain
\begin{eqnarray}
\langle \vec{\zeta}|H|\Psi \rangle 
&=&
\langle \vec{\zeta}| \left[
\int dl_1dl_2\left\{ 
-\frac{1}{2\beta^2}
\sum_ie^{\zeta_i(l_1+l_2)}\bar{\varphi}(l_1)\bar{\varphi}(l_2)
\right.\right.\nonumber
\\
& &
\hspace{1cm}
\left.\left.
-\frac{1}{2\beta^2}\sum_{i,j}e^{\zeta_il_1+\zeta_jl_2}\bar{\varphi}(l_1+l_2)
\right\}
+\sum_iU(\zeta_i)
\right]
|\Psi \rangle 
\nonumber
\\
&=&
\left[
-\frac{1}{2\beta^2}\sum_i\partial_{\zeta_i}^2
-\frac{1}{2\beta^2}\sum_{i\neq j}
\frac{\partial_{\zeta_i}-\partial_{\zeta_i}}{\zeta_i-\zeta_j}
+\sum_iU(\zeta_i)\right]
\langle \vec{\zeta}|\Psi \rangle 
\nonumber
\\
&=&
\left[
-\frac{1}{2\beta^2}\sum_i
\triangle(\vec{\zeta})^{-1}
\left(\frac{\partial}{\partial \zeta_i}\right)^2
\triangle(\vec{\zeta})
+\sum_iU(\zeta_i)
\right]
\langle \vec{\zeta}|\Psi \rangle .
\end{eqnarray}
Thus, the collective field 
Hamiltonian (\ref{collective field hamiltonian}) coincides with the 
Hamiltonian (\ref{eigenhamiltonian}) for the matrix eigenvalues. 

The state $|\Psi\rangle$ should be in the form 
\begin{equation}
|\Psi\rangle
=
F[\varphi ]|0\rangle_{\bar{\varphi}},
\label{form}
\end{equation}
where $|0\rangle_{\bar{\varphi}}$ is the eigenstate of $\bar{\varphi}$ with the 
eigenvalue $0$. Assuming that $~_\varphi \langle 0|0\rangle_{\bar{\varphi}} =1$, 
we obtain 
\begin{equation}
\langle \vec{\zeta}|\Psi\rangle 
=
F\biggl[\sum_ie^{\zeta_il}\biggr],
\end{equation}
in which form any symmetric function of $\zeta_i$ can be represented. 
This is the basic idea on which the collective field theory is constructed. 

\subsection{Bosonization}
Let us define
\begin{equation}
{\cal O}^\pm (\zeta )
=
\exp 
\left[
- 
\int_\epsilon^\infty \frac{dl}{l}e^{-(\zeta \pm i\delta )l}\varphi (l)
\right]
\exp 
\left[ 
\int_\epsilon^\infty \frac{dl}{l}e^{\zeta l}\bar{\varphi }(l)
\right].
\label{bosonization}
\end{equation}
Here, $\epsilon >0$ and $\delta >0$ are small numbers and we take the limit 
$\epsilon\rightarrow 0,~\delta\rightarrow 0$ eventually. $\epsilon$ is 
necessary to regularize the divergence at $l\sim 0$ in the integral. 
We consider $\zeta$ to be on the real axis and the integral 
$\int_\epsilon^\infty \frac{dl}{l}e^{-(\zeta +i\delta )l}\varphi (l)$ 
is supposed to have a cut on the real axis. 
$\delta$ specifies how to avoid the cut. 
We eventually consider these operators between the bra $~_\varphi \langle 0|$ 
and the ket $|0\rangle_{\bar{\varphi}}$. Thus we express ${\cal O}$ in the way 
that all the $\varphi$'s come on the left of $\bar{\varphi}$'s.  

The hermitian conjugate of ${\cal O}^\pm$ can be given in the above-mentioned 
operator ordering as 
\begin{eqnarray}
({\cal O}^\pm )^\dagger (\zeta )
&=&
\exp 
\left[-
\int_\epsilon^\infty \frac{dl}{l}e^{\zeta l}\bar{\varphi }(l)
+2\mbox{Re}\int_\epsilon^\infty \frac{dl}{l}e^{-\zeta l}\varphi (l)
\right]
\nonumber
\\
& &
\hspace{2cm}
\times 
\exp 
\left[- 
\int_\epsilon^\infty \frac{dl}{l}e^{-(\zeta \mp i\delta )l}\varphi (l)
\right]
\nonumber
\\
&=&
\exp 
\left[
\int_\epsilon^\infty \frac{dl}{l}e^{-(\zeta \pm i\delta )l}\varphi (l)
\right]
\exp 
\left[
-\int_\epsilon^\infty \frac{dl}{l}e^{\zeta l}\bar{\varphi }(l)
\right]
\nonumber
\\
& &
\hspace{2cm}
\times
\exp 
\left[ 
\frac{1}{2}
\int_\epsilon^\infty \frac{dl}{l}
\left(e^{\pm i\delta l}-e^{\mp i\delta l}\right)
\right]
\nonumber
\\
&=&
\pm i
\exp 
\left[
\int_\epsilon^\infty \frac{dl}{l}e^{-(\zeta \pm i\delta )l}\varphi (l)
\right]
\exp 
\left[
-\int_\epsilon^\infty \frac{dl}{l}e^{\zeta l}\bar{\varphi }(l)
\right]
.
\label{bosonizationdagger}
\end{eqnarray}
Therefore the hermitian conjugate of ${\cal O}^{\pm}$ can be given essentially by flipping the 
sign of the exponent, as in the usual bosonization formula. 

We would like to use ${\cal O}^\pm$ and $({\cal O}^\pm )^\dagger $ to 
define fermionic operators satisfying the canonical commutation relations. 
In order to do so, we need to calculate the commutation relations 
for these operators. 
Let us consider the product of operators 
${\cal O}^\pm (\zeta ) {\cal O}^\pm  (\zeta^\prime )$ for example. 
In order to define such a product, we first define it in the case $\zeta^\prime >\zeta$, 
and deal with the other case by the analytic continuation. This procedure 
should be considered as a variant of the radial ordering in the usual case. 
Then we can show the following identity:
\begin{eqnarray}
{\cal O}^\pm (\zeta ){\cal O}^\pm (\zeta^\prime )
&=&
\epsilon (\zeta^\prime -\zeta \pm i\delta )
\exp 
\left[
- 
\int_\epsilon^\infty \frac{dl}{l}
\left(e^{-(\zeta \pm i\delta )l}+e^{-(\zeta^\prime \pm i\delta )l}\right)
\varphi (l)
\right]
\nonumber
\\
& &
\hspace{2.5cm}
\times
\exp 
\left[ 
\int_\epsilon^\infty \frac{dl}{l}
\left(e^{\zeta l}+e^{\zeta^\prime l}\right)
\bar{\varphi }(l)
\right].
\end{eqnarray}
Here we have used the following identity,
\begin{equation}
\int_\epsilon^\infty \frac{dl}{l}e^{-\zeta l}
\sim 
-\ln (\epsilon \zeta ).
\end{equation}
We can prove similar formulas for other products and 
show 
\begin{eqnarray}
\{ {\cal O}^\pm (\zeta ), ({\cal O}^\pm )^\dagger (\zeta^\prime )\}
&=&
\frac{2\pi}{\epsilon}\delta (\zeta -\zeta^\prime )
\nonumber
\\
\{ {\cal O}^\pm (\zeta ), ({\cal O}^\mp )^\dagger (\zeta^\prime )\}
&=&
0
\nonumber
\\
\{ {\cal O}, {\cal O} \} 
&=&
0
\nonumber
\\
\{ {\cal O}^\dagger , {\cal O}^\dagger \} 
&=&
0
. 
\end{eqnarray}

Now let us define the fermionic operators
\begin{eqnarray} 
\psi (\zeta )
&\equiv &
\sqrt{\frac{\epsilon}{4\pi}}
[{\cal O}^+ +{\cal O}^-](\zeta )
\nonumber
\\
\psi^\dagger (\zeta )
&\equiv &
\sqrt{\frac{\epsilon}{4\pi}}
[({\cal O}^+)^\dagger +({\cal O}^-)^\dagger ](\zeta )
\label{psipsidagger}
\end{eqnarray}
which satisfy the canonical anti-commutation relation:
\begin{equation}
\{ \psi (\zeta ),\psi^\dagger (\zeta^\prime ) \} 
=
\delta (\zeta -\zeta ^\prime ).
\end{equation}
$\psi^\dagger $ can be considered as the creation operator 
and since 
\begin{equation}
~_\varphi \langle 0|\psi^\dagger (\zeta )=0, 
\end{equation}
$~_\varphi \langle 0|$ can be considered as the vacuum. 

We can define another pair of canonical fermions as 
\begin{eqnarray} 
\tilde{\psi}^\dagger (\zeta )
&\equiv &
\sqrt{\frac{\epsilon}{4\pi}}
[{\cal O}^+ -{\cal O}^-](\zeta )
\nonumber
\\
\tilde{\psi} (\zeta )
&\equiv &
\sqrt{\frac{\epsilon}{4\pi}}
[({\cal O}^+)^\dagger -({\cal O}^-)^\dagger ](\zeta ).
\end{eqnarray}
They satisfy 
\begin{equation}
\{ \tilde{\psi} (\zeta ),\tilde{\psi}^\dagger (\zeta^\prime ) \} 
=
\delta (\zeta -\zeta ^\prime )
\end{equation}
and 
\begin{equation}
~_\varphi \langle 0|\tilde{\psi}^\dagger (\zeta )=0, 
\end{equation}
and anti-commute with $\psi$ and $\psi^\dagger$.
Thus we have constructed fermions $\psi$ and $\tilde{\psi}$ which acts on the Hilbert space 
of collective field theory. 

\subsection{Description via fermions}
Using the fermionic operator $\psi (\zeta )$, we can express the 
inner product of the wave function $\Psi (\vec{\zeta})$ in the framework 
of the collective field theory. Noting that 
\begin{equation}
~_\varphi \langle 0|\psi (\zeta_1 )\cdots \psi (\zeta_N)
=
\langle \vec{\zeta}|\triangle (\vec{\zeta})
\epsilon^{\frac{N^2}{2}}
(\pi )^{-\frac{N}{2}},
\end{equation}
we can express 
the fermion wave function defined by 
$\Psi_{\rm fermion}(\vec{\zeta})\equiv \triangle (\vec{\zeta})\Psi (\vec{\zeta})$
as 
\begin{equation}
\Psi_{fermion}(\vec{\zeta})
=
\epsilon^{-\frac{N^2}{2}}\pi^{\frac{N}{2}}
~_\varphi \langle 0|
\psi (\zeta_1 )\cdots \psi (\zeta_N)
|\Psi \rangle .
\end{equation}
This relation is exactly the one between the second quantized fermion 
operator and the many body wave function. 
The partition function of the matrix quantum mechanics is expressed as 
\begin{equation}
Z
\propto
\int \prod_id\zeta_i 
\langle \Psi |
\psi^\dagger (\zeta_N)\cdots \psi^\dagger (\zeta_1 )
|0\rangle_\varphi 
~_\varphi \langle 0|
\psi (\zeta_1 )\cdots \psi (\zeta_N)e^{-iHT}
|\Psi \rangle ,
\end{equation}
where the state $|\Psi\rangle$ corresponds to the boundary condition. 
For a large $T$, $Z\sim e^{-iE_0T}$ and $E_0$ can be identified with the free energy 
of the string theory. 
Since one can show 
\begin{eqnarray}
[H,\psi (\zeta )]
&=&
- \left(-\frac{1}{2\beta^2}\partial_\zeta^2+U(\zeta )\right)
\psi (\zeta ),
\nonumber
\\
\mbox{[}H,\psi^\dagger (\zeta )]
&=&
 \left(-\frac{1}{2\beta^2}\partial_\zeta^2+U(\zeta )\right)
\psi^\dagger (\zeta ),
\end{eqnarray}
and $~_\varphi \langle 0|H=0$, $E_0$ is the lowest energy eigenvalue of the 
$N$ fermion state. 
Thus the fermion operator $\psi (\zeta )$ is exactly the 
nonrelativistic free fermion which describes the matrix quantum mechanics. 

The other fermions $\tilde{\psi},\tilde{\psi}^\dagger$ correspond 
to matrix eigenvalues with ''negative norm''. 
Indeed 
\begin{equation}
~_\varphi \langle 0|\tilde{\psi}(\zeta_1)\cdots\tilde{\psi}(\zeta_N)\varphi(l)
=
~_\varphi \langle 0|\tilde{\psi}(\zeta_1)\cdots\tilde{\psi}(\zeta_N)
\left(-\sum_ie^{\zeta_i l}\right).
\end{equation}
Such eigenvalues appear in the supermatrix model. Therefore the collective 
field theory can express the super matrix model. 
It is natural for the collective field Hilbert space to include such modes, 
because it includes $\pi (\zeta )$ conjugate to $\rho (\zeta )$, which makes 
it impossible for $\rho (\zeta )$ to take only nonnegative values. 
Moreover since
\begin{eqnarray}
[H,\tilde{\psi} (\zeta )]
&=&
\left(-\frac{1}{2\beta^2}\partial_\zeta^2+U(\zeta )\right)
\tilde{\psi} (\zeta ),
\nonumber
\\
\mbox{[}H,\tilde{\psi}^\dagger (\zeta )]
&=&
 -\left(-\frac{1}{2\beta^2}\partial_\zeta^2+U(\zeta )\right)
\tilde{\psi}^\dagger (\zeta ),
\end{eqnarray}
if we take the potential $U(\zeta )$ to be bounded below, the fermion 
$\tilde{\psi}^\dagger$ generate the energy spectrum not bounded below. 

\section{The double scaling limit}
So far we have been studying the matrix quantum mechanics without taking the 
continuum limit. 
Now let us take the double scaling limit of the bosonization rule given above. 
Putting $U(\zeta )=\frac{1}{2}\zeta^2$, we take the limit $\beta\rightarrow\infty$. 
The continuum string field should be defined as 
\begin{equation}
\varphi_c (l)\equiv \varphi (\sqrt{\beta}l).
\end{equation}
We use the variable $y=\sqrt{\beta}\zeta$ to describe the continuum variables. 
The Hamiltonian in the continuum limit is given as 
$H_c\equiv \beta H$. 
Introducing the chemical potential $\mu$, we get the 
continuum Hamiltonian as 
\begin{eqnarray}
H_c
&=&
-\frac{1}{2}
\int dl_1dl_2
\left[\varphi_c (l_1)\varphi_c (l_2)\bar{\varphi}_c(l_1+l_2)
+\varphi_c (l_1+l_2)\bar{\varphi}_c(l_1)\bar{\varphi}_c(l_2)\right]\nonumber
\\
& &
\hspace{0.5cm}
+\int dl\varphi_c (l)
\left(-\frac{1}{2}\delta^{\prime\prime} (l)
+\mu \delta (l)\right).
\label{continuumHamiltonian}
\end{eqnarray}

${\cal O}^\pm$ can be rewritten in terms of the continuum variables as 
\begin{eqnarray}
{\cal O}^\pm (\zeta )
&=&
\exp 
\left[- 
\int_\epsilon^\infty \frac{dl}{l}e^{-(\zeta \pm i\delta )l}\varphi (l)
\right]
\exp 
\left[
\int_\epsilon^\infty \frac{dl}{l}e^{\zeta l}\bar{\varphi }(l)
\right]
\nonumber
\\
&=&
\exp 
\left[
- 
\int_{\frac{\epsilon}{\sqrt{\beta}}}^\infty 
\frac{dl}{l}e^{-(y \pm i\sqrt{\beta}\delta )l}\varphi_c (l)
\right]
\exp 
\left[ 
\int_{\frac{\epsilon}{\sqrt{\beta}}}^\infty 
\frac{dl}{l}e^{y l}\bar{\varphi }_c(l)
\right].
\end{eqnarray}
Thus by replacing $\zeta$ by $y$ and rescaling $\epsilon$ and $\delta$ 
by $\sqrt{\beta}$, all the bosonization rules are the same 
as the ones given in the previous section, and $\psi ,\psi^\dagger$ 
are defined accordingly.  
We will omit the subscript $c$ and forget that $\epsilon$ and $\delta$ are 
rescaled by $\sqrt{\beta}$ in the following. 
The continuum Hamiltonian is given in terms of the fermions as 
\begin{equation}
\int dy
\left[\psi^\dagger (y )
\left(-\frac{1}{2}\partial_{y}^2
 -\frac{1}{2}y^2+\mu \right)\psi (y )
-\left(-\frac{1}{2}\partial_{y}^2
 -\frac{1}{2}y^2+\mu \right)
\tilde{\psi}^\dagger (y )\tilde{\psi}(y)\right]. 
\end{equation}

The continuum limit of the Das-Jevicki variables are defined 
in the same way. 
From the relation (\ref{rho phi}), we can get 
\begin{equation}
\varphi (y)
=
\int dy^\prime \frac{\rho (y^\prime )}{y-y^\prime },
\label{relation 1}
\end{equation}
and the relation (\ref{phi p}) implies 
\begin{equation}
\varphi (y\pm i\delta )+\bar{\varphi}(-y)
=
i\partial_y\pi (y)
\mp i\pi \rho (y),
\label{relation 2}
\end{equation}
where
\begin{eqnarray}
\varphi (y )
&\equiv&
\int_0^\infty dle^{-yl}\bar{\varphi}(l),
\nonumber
\\
\bar{\varphi}(-y)
&\equiv&
\int_0^\infty dle^{yl}\bar{\varphi}(l),
\end{eqnarray}
which gives the relation between the variables in the continuum limit. 
The collective Hamiltonian in the Das-Jevicki form becomes
\begin{equation}
H_c
=
\int dy
\left[-\frac{1}{2}(y^2-2\mu )\rho (y)
+\frac{1}{2}(\partial_y \pi (y))^2\rho (y)
+\frac{\pi^2}{6}\rho^3(y)\right].
\label{continuum DJ}
\end{equation}

\section{Perturbative calculations}
Since we have the exact expression for the fermion variables in terms 
of the string field, in principle, we can calculate the amplitudes involving fermions 
perturbatively using the string field. 
In this section, we perform some calculations for a simple example, 
and point out a subtlety involved in such calculations. 

\subsection{Expansions of the Das-Jevicki variables}
In order to consider the theory around the vacuum, 
Das-Jevicki variables are more convenient than $\varphi ,\bar{\varphi}$. 
From the continuum Hamiltonian (\ref{continuum DJ}), we can see that 
the following distribution yields
a static vacuum configuration:
\begin{equation}
\rho_0(y)=
\begin{cases}
\displaystyle{\frac{1}{\pi}}
\sqrt{y^2-2\mu}& \textrm{for}\quad y
\leq  
-\sqrt{2\mu} \\
0 & \textrm{for}\quad y>-\sqrt{2\mu}
\end{cases}.
\end{equation}
The collective field theory around this vacuum is most conveniently 
described by introducing the variable $\tau$ which satisfies 
\begin{equation}
y=-\sqrt{2\mu}\cosh \tau ,
\end{equation}
and the field $\phi (\tau )$ and its canonical conjugate $\pi_\phi (\tau )$ 
as
\begin{eqnarray}
\rho (y)
&=&
\frac{1}{\pi}\sqrt{y^2-2\mu}
+\frac{1}{\sqrt{\pi}}
\frac{\partial_{\tau}\phi (\tau )}{\sqrt{y^2-2\mu}}
\nonumber
\\
\partial_y\pi (y)
&=&
\sqrt{\pi}
\frac{\pi_\phi (\tau )}{\sqrt{y^2-2\mu}},
\label{pi phi}
\end{eqnarray}
or 
\begin{eqnarray}
ip_\mp (y)
&\equiv&
i\partial_y\pi (y)\mp i\pi \rho (y)
\nonumber
\\
&=&
i\left[\mp \sqrt{y^2-2\mu}
+\frac{\sqrt{\pi}}{\sqrt{y^2-2\mu}}
(\pi_\phi (\tau )\mp \partial_\tau \phi (\tau ))\right].
\label{relation 3}
\end{eqnarray}
Using these variables, the Hamiltonian for the fluctuation becomes 
\begin{equation}
H_c
=
\int_0^\infty d\tau 
\left[\frac{1}{2}(\partial_\tau \phi )^2+
\frac{1}{2}(\pi_\phi )^2+
\frac{\sqrt{\pi}}{4\mu \sinh^2\tau}(\pi_\phi )^2\partial_\tau \phi +
\frac{\sqrt{\pi}}{12\mu \sinh^2\tau}(\partial_\tau \phi )^3\right].
\label{hamiltonian phi}
\end{equation}
Now the Hamiltonian becomes the one for the massless boson $\phi $ in the 
two dimensional spacetime $(t,\tau )$ with nonrelativistic interactions. 
One can expand the operators $\phi $ and $\pi_\phi$ as 
\begin{eqnarray}
\phi (\tau )
&=&
\int_0^\infty \frac{dE}{\sqrt{\pi E}}\left(b(E)+b^\dagger (E)\right)\sin (E\tau ),
\nonumber
\\
\pi_\phi (\tau )
&=&
i\int_0^\infty dE\sqrt{\frac{E}{\pi}}\left(b^\dagger (E)-b(E)\right)\sin (E\tau ),
\label{b bdagger}
\end{eqnarray}
where the boundary conditions for $\phi$ and $\pi_\phi$ are chosen as \cite{Das-Jevicki} 
\begin{equation}
\phi (\tau )|_{\tau =0}=\pi_\phi (\tau )|_{\tau =0}=0. \label{bc}
\end{equation}
$b$ and $b^\dagger$ are hermitian conjugate to each other and satisfies 
\begin{equation}
[b(E),b^\dagger (E^\prime )]=\delta (E-E^\prime ). 
\end{equation}
The free part of the Hamiltonian (\ref{hamiltonian phi}) becomes 
\begin{equation}
\int_0^\infty dE E b^\dagger (E)b(E). 
\label{diagonal}
\end{equation}

\subsection{Perturbative calculations}
Substituting eq.(\ref{b bdagger}) into eq.(\ref{relation 3}) 
and using eq.(\ref{relation 2}), 
we can expand the fields $\varphi ,\bar{\varphi}$ in terms of the 
oscillators $b(E), b^\dagger (E)$. 
In principle it is possible to perform perturbative calculations 
using these oscillators. 

Here let us study the fermionic operators taking the one loop 
effects into account.  
Using eq.(\ref{bosonization}), 
it is straightforward to see that for $y\leq -\sqrt{2\mu}$ 
the fermions can be written as 
\begin{eqnarray}
{\cal O}^\pm (y)
&=&
\exp 
\left[
\int_\Lambda^{y\pm i\delta}dy^\prime \varphi (y^\prime )
\right]
\exp 
\left[ 
\int_{-\Lambda}^ydy^\prime \bar{\varphi }(-y^\prime )
\right]
\nonumber
\\
&=&
\exp \left(
\mp i\int^y_{-\sqrt{2\mu}} dy^\prime \sqrt{y^{\prime 2}-2\mu}
\right)
\nonumber
\\
& &
\times
\exp
\left[
-
\int_0^\infty dE\left(b(E)+b^\dagger (E)\right)
\frac{\cos E(\tau \pm \pi i)-\cos E\lambda}{\sqrt{E}\sinh \pi E}
\right]
\nonumber
\\
& &
\times
\exp
\left[
\int_0^\infty dE\left(e^{\pi E}b(E)+e^{-\pi E}b^\dagger (E)\right)
\frac{\cos E\tau -\cos E\lambda}{\sqrt{E}\sinh \pi E}
\right].
\label{calOpert}
\end{eqnarray}
We have introduced $\Lambda \sim \epsilon^{-1}$ to regularize the 
integral in place of $\epsilon$,\footnote{
We have assumed that 
$\Lambda$ is much bigger than any $y$ such that 
$\rho_0(y)\neq 0$. 
This is justified because our bosonization rule is 
valid even before the continuum limit. 
}
 and we set
\begin{eqnarray}
y
&=&
-\sqrt{2\mu}\cosh \tau ,
\nonumber
\\
\Lambda 
&=&
\sqrt{2\mu}\cosh \lambda . 
\end{eqnarray}
Rewriting eq.(\ref{calOpert}) in the normal ordered form with respect to 
$b,b^\dagger$, we can take the one-loop effects into account. 
We obtain
\begin{eqnarray}
{\cal O}^\pm (y)
&=& C
\Lambda^{\frac{1}{2}}(\ln \Lambda )^{-\frac{3}{4}}
(y^2-2\mu )^{-\frac{1}{4}}
\exp 
\left(
\mp i \int^y_{-\sqrt{2\mu}} dy^\prime \sqrt{y^{\prime 2}-2\mu}
\right)
\nonumber
\\
& &
\times
\exp 
\left(
-\int_0^\infty \frac{dE}{\sqrt{E}}b^\dagger (E)e^{\pm iE\tau}
\right)
\exp 
\left(
\int_0^\infty \frac{dE}{\sqrt{E}}b (E)e^{\mp iE\tau}
\right),
\label{divergences}
\end{eqnarray}
where $C$ denotes a numerical constant.
This form of the fermionic operator is quite like those found 
in \cite{Gross-Klebanov}. 
Especially we get the WKB wave function precisely as a factor. 
However it is with a divergent constant. 

Actually such a calculation is subtle for $y\leq -\sqrt{2\mu}$. 
In order to deal with the matrix model, we should restrict ourselves to the 
states in the string field Hilbert space, 
which are annihilated by $\tilde{\psi}^\dagger$. 
Suppose we are given 
a coherent state $\langle \varphi_0|$ which satisfy
\begin{equation}
\langle \varphi_0|\varphi (l)
=
\langle \varphi_0|\varphi_0 (l),
\end{equation}
let us examine what conditions $\langle \varphi_0|$ should satisfy in order 
to be a state annihilated by $\tilde{\psi}^\dagger$. 
We assume that $\varphi_0 (l)$ can be written as 
\begin{equation}
\varphi_0 (l)
=
\int d\zeta e^{\zeta l}\rho_0 (\zeta ),
\end{equation}
with $\rho_0(\zeta )\geq 0$. Since 
\begin{eqnarray}
\langle \varphi_0|\tilde{\psi}^\dagger (\zeta )
&=&
\langle \varphi_0|
\sqrt{\frac{\epsilon}{4\pi}}
[{\cal O}^+ -{\cal O}^-](\zeta )
\nonumber
\\
&\propto&
\langle \varphi_0|
\left\{
\exp 
\left[- 
\int_\epsilon^\infty \frac{dl}{l}e^{-(\zeta + i\delta )l}\varphi (l)
\right]
-
\exp 
\left[- 
\int_\epsilon^\infty \frac{dl}{l}e^{-(\zeta - i\delta )l}\varphi (l)
\right]\right\}
\nonumber
\\
& &
\hspace{5mm}
\times
\exp 
\left[ 
\int_\epsilon^\infty \frac{dl}{l}e^{\zeta l}\bar{\varphi }(l)
\right]
\nonumber
\\
&\propto&
\langle \varphi_0|
\left\{
 \exp
 \left[
  \int d\zeta^\prime 
  \ln \left[\epsilon (\zeta -\zeta^\prime +i\delta )\right]
  \rho_0 (\zeta^\prime )
 \right]
\right.
\nonumber\\
&&
\left.
 \qquad\qquad 
 -
 \exp
 \left[
  \int d\zeta^\prime 
  \ln \left[\epsilon (\zeta -\zeta^\prime -i\delta )\right]
  \rho_0 (\zeta^\prime )
 \right]
\right\},
\end{eqnarray}
$\langle \varphi_0|\tilde{\psi}^\dagger (\zeta )=0$ when the two terms  in 
the last line cancel with each other. When 
$\rho_0 (\zeta )\geq 0$, we expect that the difference between the two 
comes from the difference in the imaginary parts of the exponents which is 
\begin{equation}
2\pi i \int^\zeta d\zeta^\prime \rho_0 (\zeta^\prime ).
\end{equation}
Therefore if and only if $\int^\zeta d\zeta^\prime \rho_0 (\zeta^\prime )$ 
is an integer, 
$\langle \varphi_0|\tilde{\psi}^\dagger (\zeta )=0$. 
This is automatically satisfied before the continuum limit is taken, but 
it is a very subtle condition in the continuum limit. 
In the vacuum configuration, eigenvalues are distributed in the region 
$y\leq -\sqrt{2\mu}$. 
Hence for $y\leq -\sqrt{2\mu}$, this condition is very subtle. 
The divergences in eq.(\ref{divergences}) are the signs of this subtlety. 

The situation is quite similar to that in the $c=0$\cite{Ishibashi:2005dh} 
and $c<1$\cite{Ishibashi:2005zf} case.
In those cases, we encounter divergences similar to the ones in 
eq.(\ref{divergences}). However, in calculating physical quantities such as 
the chemical potential of instantons, 
they are cancelled by divergences from other factors and we eventually obtain 
finite results. 
We expect that similar things happen in $c=1$ case.

\section{Conclusions and discussions}
In this paper, we construct a string field theory for $c=1$ noncritical string theory, 
using the loop variables. 
We give an exact bosonization rule, by which we can express the nonrelativistic 
free fermions in terms of the string field. 
The description by the string field involves fermions with negative norm and energy 
besides the usual fermion. 
This is inevitable because we should introduce a canonical conjugate to the 
eigenvalue distribution function $\rho$ in the string field theory. 
The existence of such extra fermions causes subtleties in the perturbative calculations. 

We argue that the Hamiltonian written in terms of the loop variables 
is equivalent to the classical part of Das-Jevicki's Hamiltonian. 
Higher order terms are necessary for Das-Jevicki's Hamiltonian 
to reproduce the results of the matrix quantum mechanics. 
On the other hand, the 
nonrelativistic fermion formulation of the matrix quantum mechanics 
is reproduced from the string field theory. 
It is conceivable that our prescription of normal ordering of the fermionic operators 
has something to do with this discrepancy. 
It is an important and intriguing issue to clarify this point by 
comparing our results with 
the perturbative calculations in 
\cite{Demeterfi:1991tz}\cite{Demeterfi:1991nw}, for example. 
We leave it as a future problem. 

In \cite{Alexandrov:2004cg}, nonperturbative effects in $c=1$ string theory
were investigated using another bosonization rule.
The authors in \cite{Alexandrov:2004cg} showed 
that there is a nonperturbative 
 correction 
on the zero mode of bosonic fields. 
In our rule, on the other hand,  
the bosons $\phi(\tau)$ and $\pi_\phi(\tau)$ cannot have any zero mode,
by construction,
due to the boundary condition (\ref{bc}).
This suggests that their approach and ours will give different results 
for nonperturbative effects.

Since the fermions in $c=1$ strings can be considered as D-branes, 
our results will be useful in understanding how one should consider D-branes 
in the context of closed string field theory and clarify the relation between 
string theory and the matrix models. 
The Hamiltonian for our string field involves joining-splitting interaction 
which is similar to that of the light-cone gauge string field theory for critical 
strings. 
Therefore we may be able to use our approach in the critical string theory 
to investigate the above mentioned issues. 

\acknowledgments

We would like to thank H.~Kawai and I.K.~Kostov for discussions.

\end{document}